\documentclass{elsart}
\usepackage{amssymb}
\usepackage{epsfig,epsf,graphicx}
\usepackage{amsmath}

\def\nl{\noindent}

\def\sla#1{\rlap\slash #1}

\newcommand{\be}{\begin{equation}}
\newcommand{\ee}{\end{equation}}

\newcommand{\bee}{\begin{eqnarray}}
\newcommand{\eee}{\end{eqnarray}}

\def\be{\begin{eqnarray} &&}

\def\ee{\end{eqnarray}}
\def\bew{\begin{widetext}}
\def\ew{\end{widetext}}

\begin{document}
\begin{frontmatter}
\title{Light-Front projection of spin-1 electromagnetic
current and zero-modes}
\author{J. P. B. C. de Melo$^a$ and T. Frederico$^b$}
\address[LFTC]{Laborat\'orio de F\'\i sica Te\'orica e Computa\c{c}\~ao
Cient\'\i fica - LFTC \\
Universidade Cruzeiro do Sul, 01506-000, S\~ao Paulo, SP, Brazil}
\address[ITA]{Instituto Tecnol\'ogico de
Aeron\'autica, DCTA \\ 12.228-900 S\~ao Jos\'e dos Campos, SP,
Brazil.}

\date{\today}
\maketitle
\begin{abstract}
The issue of the contribution of zero-modes to the light-front
projection of the electromagnetic current of phenomenological models
of vector particles vertices is addressed in the Drell-Yan frame.
Our analytical model of the Bethe-Salpeter amplitude of a spin-1
fermion-antifermion composite state gives a physically motivated
light-front wave function symmetric by the exchange of the fermion
and antifermion, as in the $\rho$-meson case. We found that among
the four independent matrix elements of the plus component in the
light-front helicity basis only the $0\to 0$ one carries zero mode
contributions. Our derivation generalizes to symmetric models,
important for applications, the above conclusion found  for a
simplified non-symmetrical form of the spin-1 Bethe-Salpeter
amplitude with photon-fermion point-like coupling and also for a
smeared fermion-photon vertex model.
\end{abstract}
\end{frontmatter}

The electromagnetic from factors of composite spin-1 particles as
the $\rho$-meson,  have been addressed with increasing interest in
the last years (see e.g.
\cite{Carda95,Pacheco97,JI2002,Jaus2003,Choi2004,Adamuscin2007,Grigoryan2007,Aliev2009,Garcia2010,Sang2010,Roberts2011,Choi2011}),
being an instrument to investigate the hadronic structure in terms
of their basic constituents. Particularly Light-Front (LF) models
are useful to describe the composite structure of hadrons in terms of
constituent quarks degrees of freedom, which
despite the inherent simplicity implement correctly kinematical
boost properties of the corresponding
amplitudes in exclusive processes~\cite{Terentev76,Brodsky98}.

However, the light-front description of a physical state in a
truncated Fock-space basis breaks the rotational symmetry, as the
associated transformation corresponds to a dynamical
boost~\cite{Pacheco97,Pacheco98,Naus98,Pacheco992}. It is a
formidable task to study the transformation properties of the
Fock-space wave function under dynamical boosts in light-front
quantization~\cite{Brodsky98}. An analysis starting with covariant
and analytical models of the Bethe-Salpeter  (BS) amplitude   are
helpful to pin down the missing features in respect to boosts
transformations in a truncated LF Fock-space description of the
composite system. The projection onto the LF of a field theoretical
model BS equation can be performed by integrating the relevant
loop-integrals over the LF energy $k^-=k^0-k^3$. It can be done
systematically trough a quasi-potential technique \cite{sales2000}
allowing to define concomitantly the relevant operators to be used
with the valence wave function \cite{Marinho2008} (see also
\cite{FS2011,Fre2011}).

The lack of contributions of higher Fock-components beyond the
valence wave function for spin-1 composite systems, breaks the rotational symmetry relations
between the matrix elements of the plus component of the
electromagnetic current ($J^+=J^0+J^3$) in the Drell-Yan frame
(momentum transfer $q^+=q^0+q^3=0$). For example, this was verified by
starting with a
covariant model of the $\rho$-meson as a $q\bar q$ bound
state~\cite{Pacheco97,Pacheco98}. It was shown that, if pair-term
(Z-diagram) contributions are ignored in the evaluation of the
matrix elements of the electromagnetic current, the covariance of
the form factors is
lost~\cite{Pacheco97,JI2002,Choi2004,Pacheco98,Pacheco992}. Pair terms appearing in
the matrix elements of current operator are associated with nonvalence contributions
(see e.g. \cite{Marinho2008}).

Within a field theoretic framework, the nonvalence terms in the Mandelstam formula or
impulse approximation are due to the coupling between the valence and higher Fock sectors
in the LF hamiltonian. In this respect, the nonvalence terms can be translated to two-body
current operators  acting on the valence sector (see e.g. \cite{Marinho2008,HK2004}).
The covariance of the form factors is broken if nonvalence contributions to the current
are disregarded and only valence matrix elements are computed. In the
Drell-Yan frame, the surviving pair diagrams are associated with
zero modes in the limit of $q^+\to 0$. It remains an open question, if
the singular behavior at the end points of naive analytical
covariant vertex models,  which originates zero-modes in the
computation of exclusive process in the Drell-Yan frame, are indeed
brought by light-front QCD, or by hadronic models with meson
exchange currents projected onto the light-front \cite{Marinho2008}.

The problem of the missing covariant properties of the form factors
computed with the LF valence wave function, motivated different
approaches to obtain the associated observables.
This is the case of the explicitly covariant Light-Front dynamics
\cite{karmanovprp} or the introduction of other kinematical conditions together with
a Poincar\'e Covariant form of the current operator as done by
Lev, Pace and Salm\`e (LPS) \cite{LevPaceSalme}.  The form factors in the
LPS framework are computed  with momentum transfer along the
z-direction in the Breit-Frame. Noteworthy to observe that, the Z-diagram gives an
important part of the form factor of strongly bound
systems, as the pion, with the choice of momentum transfer
according to LPS  (see e.g. \cite{Bakker2001,Pacheco2002}).

For spin-1 particles, the missing rotational properties of the microscopic
matrix elements of $J^+$ computed only with valence states in the Breit-frame with $q^+=0$,
implies in an ambiguity in the extraction of the form factors~\cite{Inna84,Inna89,Karmanov96}.
In this case $J^+$ has four independent matrix elements, although only three form
factors exist due to the constraints of covariance and current
conservation. The matrix elements are related by an
identity, namely, the angular condition (see e.g. \cite{Inna84}), which is
violated by computing matrix elements of one-body current
operators only with the valence  component of the LF wave function.
In this regard, several extraction schemes for evaluating spin-1 form factors
were proposed \cite{Inna84,CKP1988,BH1992,Frankfurt93}.

The different extraction schemes were tested through the calculation of the
$\rho$-meson electromagnetic form factors, with a covariant model of
the quark-antiquark-$\rho$ vertex, both in a covariant form and
using the LF projection to the valence sector \cite{Pacheco97}. The
prescription proposed by Grach and Kondratyuk (GK) \cite{Inna84} provides
form factors in agreement with the fully covariant calculation.
Later on, Ref. \cite{JI2002} demonstrated that the above prescription
eliminates the nonvalence zero-mode contributions to the form factors computed
with a smeared photon vertex model.
In view of the interest in exclusive processes involving composite spin-1 particles,
like the $\rho-$meson or the deuteron, it remains the question how this
analytical conclusion generalizes to different models of the
spin-1 BS amplitude, which have a LF valence wave function with a
structure beyond the asymptotic one.

Here, we study an analytical BS amplitude model of the spin-1
fermion-antifermion composite state, which has a physically
motivated valence wave function, symmetric by the exchange of the
fermion and antifermion as required by phenomenological
$\rho$-meson LF models (see e.g.\cite{Jaus2003}). We show analytically that
zero-mode contributions appearing in the matrix elements of the spin-1
electromagnetic current are canceled in the computation of form factors with
the prescription suggested by Ref.~\cite{Inna84}.

The present results goes beyond previous findings and establish for symmetric vertex models,
appealing for phenomenological applications, that the matrix element of $J^+$ computed
in the LF helicity basis are free from zero-modes apart
the $0\to 0$ one, as has been previously
found for a simplified non-symmetrical form of the $\rho$-meson
vertex in \cite{Pacheco97} and  for a smeared photon vertex model in
\cite{JI2002}. Our analysis starts with the instant form (IF)
polarization basis in cartesian representation,
because the angular condition has an intuitive form, relating the
diagonal matrix elements of $J^+$ with polarization states
transverse to the momentum transfer\cite{Frankfurt93}. The cartesian
basis permits to isolate in a transparent form the zero-modes in all
matrix elements, adding further insight to the calculations with the
helicity basis in the Breit-frame for $q^+=0$.

We should add that, the pion form factor obtained from $J^+$ does
not have contributions from pair terms in the Breit-frame and
$q^+=0$, as found for an analytical and covariant pion vertex model
with a $\gamma^5$ coupling to the quarks
\cite{Pacheco2002,Pacheco99}. It is enough to evaluate the valence
part of the matrix element of the current to reproduce the covariant
result. However, a zero-mode contributes to the matrix element of
the current $J^-=J^0-J^3$ as shown in \cite{Pacheco99}. Another
example of the contribution of light-front zero-modes to the
electromagnetic form factors and their cancellation was given in
\cite{Karmanov07} in the case of a spin 1/2 fermion.

{\it Symmetrical light-front model for vector particles.} The adopted spin
and momentum structure of the $\rho-q\bar{q}$ vertex, as a prototype
of spin-1 composite particle, is given by a symmetrical expression with constituent
fermions of mass $m$. It comes as a generalization of
the spin one composite particle vertex proposed in ~\cite{Pacheco97}
and the one given in ~\cite{Pacheco2002} used for the evaluation of the pion form
factor. Our model for the vector-particle vertex reads:
\begin{equation}
\Lambda^{\mu} (k,p)  = \gamma^{\mu} \left[D(k^\prime)\right]^{-2}
-\frac{m_{v}}{2} \left( k^\mu + k^{\prime \mu}\right)\left[D_v( k)
D^2(k^{\prime})\right]^{-1}+\left[ k\leftrightarrow-k^\prime \right]
~, \label{newsymm}
\end{equation}
where $m_v$ is the mass of the vector particle,
$D(k)=(k^2 - m^2_R+ \imath \epsilon)$, $D_v(k)=(p . k+m_{v} m
- \imath \epsilon)$ and $k^{\prime}=k-p$. The regularization
function is enough to render finite the photo-absorption amplitude.
The regularization parameter is $m_R$.

The valence component of the light-front wave function is the
projection at $x^+=0$ of the Bethe-Salpeter amplitude with the
instantaneous terms in the external fermionic legs separated
out~\cite{sales2000}. Only the LF time propagating part
of the Dirac propagator is left in the external legs.
The resulting valence wave function with the vector-particle
polarized along $\epsilon_{i}$ is given by:
\begin{equation}
\Phi^{(i)}_{LF}(k^+,\vec k_\perp;p)   = \int \frac{dk^-}{2\pi
\imath} \frac{(\sla{k}_{on}+m)} {(k^2-m^2+\imath \epsilon)}\;
\epsilon_{i}\cdot\Lambda(k,p) \;
\frac{(\sla{k}^\prime_{on}+m)} {(k^{\prime 2}-m^2+ \imath
\epsilon)} \ ,
\end{equation}
where $k^\prime=k-p$. After integration over $k^-$, we get:
\begin{equation}
\Phi_{LF}^{(i)}(k^+,\vec k_\perp;p)   = \frac{\sla{k}_{on}+m}{  k^+}
~ \frac{\Lambda^{(i)}_{LF}(k^+,\vec k_\perp;p)}{ p^--p^-_0} ~
\frac{\sla{k}^\prime_{on}+m }{p^+-k^+ } \ ,\end{equation} where the
LF momenta are $k^\pm=k^0\pm k^3$ and
$k_{\perp}=(k_x,k_y)$). The support of the LF wave function is
$0<k^+< p^+$ and $p^-_0$ is the minus component of the free momentum
of the quark-antiquark system. The minus-on-shell momentum $k_{on}$
and $k^\prime_{on}$ have minus components $k^-_{on}=(\vec
{k}_\perp^2+m^2)/k^+$ and $k^{\prime -}_{on}=((\vec {p}-\vec
{k})_\perp^2+m^2)/(k^+-p^+)$, respectively. The momentum part of the
LF vertex is
\begin{small}
\begin{equation}
\Lambda^{(i)}_{LF}(k^+,\vec k_\perp;p)=
\frac{1}{((p-k_{on})^2-m^2_R)^2} \left\{ \sla{\epsilon}_i
-\frac{m_{v}}{2}
\frac{\epsilon_i\cdot\left( 2k_{on} -p\right)} {(p.k_{on}+m_v m)}\right\}
+ \left[ k \leftrightarrow -k^\prime \right]  \ , ~~~~~~ \label{lfwfrho}
\end{equation}
\end{small}
which is symmetrical under the exchange between the quark and antiquark.

{\it Electromagnetic current in impulse approximation and notation.}
The electromagnetic current for spin-1 particles has the
following general form (see e.g. \cite{Frankfurt79}):
\begin{equation}
J_{\alpha \beta}^{\mu}=[F_1(q^2)g_{\alpha \beta} -F_2(q^2)
\frac{q_{\alpha}q_{\beta}}{2 m_v^2}] (p^\mu + p^{\prime \mu})
 - F_3(q^2)
(q_\alpha g_\beta^\mu- q_\beta g_\alpha^\mu) \ , \label{eq:curr1}
\end{equation}
where $q^\mu$ is the
momentum transfer, $p^\mu$ and $p^{\prime  \mu}$ is on-shell initial
and final momentum respectively. From the covariant form factors
$F_1$, $F_2$ and $F_3$, one can obtain the charge ($G_0$),
magnetic ($G_1$) and quadrupole ($G_2$) form factors (see
e.g.~\cite{Pacheco97}).

The matrix elements of the electromagnetic current
${\mathcal J}_{ji}={\epsilon^\prime_j}^\alpha \epsilon^\beta_iJ_{\alpha
\beta}^{\mu}$ in the impulse approximation are written
as~\cite{Pacheco97}:
\begin{equation}
{\mathcal J}_{ji}^{\mu}= \int [d^4k] \frac{
Tr\left[(\sla{k}+m) \Lambda_\alpha(k,p^\prime) {\epsilon^\prime
_j}^\alpha (\sla{k}-\sla{p}^{\prime}+m)\gamma^{\mu}
(\sla{k}-\sla{p}+m) \Lambda_\beta(k,p) \epsilon^\beta_i
 \right] }{ (k^2-m+\imath
\epsilon) ((p-k)^2-m+\imath \epsilon) ((p^{\prime}-k)^2-m+\imath
\epsilon) } \ , \label{newcu}
\end{equation}
where $[d^4k]=d^4k/(2\pi)^4$, ${\epsilon^\prime_j}$ and
${\epsilon_i}$ are the polarization four-vectors of the final and
initial states, respectively.

The electromagnetic form-factors are calculated in the Breit-frame
with the Drell-Yan condition, which gives the momentum transfer
$q^\mu=(0,q_x,0,0)$. The particle initial momentum is
$p^\mu=(p^0,-q_x/2,0,0)$ and the final one is
$p^{\prime\mu}=(p^0,q_x/2,0,0)$. We use $\eta=-q^2/4 m^2_{v}$ and
$p^0=m_{v}\sqrt{1+\eta}$. The polarization four-vectors in the
instant-form basis are given by
\begin{equation}
\epsilon^{\mu}_x =(-\sqrt{\eta},\sqrt{1+\eta},0,0),~~
\epsilon^{\mu}_y=(0,0,1,0),~~  \epsilon^{\mu}_z=(0,0,0,1),
\label{inipol}
\end{equation}
for the initial state and by,
\begin{equation}
\epsilon^{\prime \mu}_x=(\sqrt{\eta},\sqrt{1+\eta},0,0),~~
\epsilon^{\prime\mu}_y=\epsilon^{\mu}_y ,~~
\epsilon^{\prime\mu}_z=\epsilon^{\mu}_z,
\label{finpol}
\end{equation}
for the final state.

{\it LF spin basis matrix elements and the angular condition.}
The matrix elements of $J^+$  in the instant-form spin basis, are
related to the matrix elements in the LF spin basis, with the unitary transformation
between these spin basis given by the Melosh rotation~\cite{Keister91}. For notational
convenience, we use $I^+$, to express the matrix elements in the LF
spin basis.   The relations between the
current matrix elements in the two spin basis are (see also
~\cite{Frankfurt93}):
\begin{small}
\begin{equation}
I^{+}_{11} =  \frac{J^{+}_{xx}+(1+\eta) J^{+}_{yy}- \eta
J^{+}_{zz} -  2 \sqrt{ \eta} J^{+}_{zx}}{2 (1+\eta)} \;,\;
I^{+}_{10}  =  \frac{\sqrt{ \eta} J^{+}_{xx} + \sqrt{\eta}
J^{+}_{zz} - (\eta-1) J^{+}_{zx}}{\sqrt{2}(1+\eta)} \ ,\nonumber
\end{equation}
\begin{equation}
 I^{+}_{1-1}  =  \frac{ (1+\eta) J^{+}_{yy}-J^{+}_{xx}+ \eta
J^{+}_{zz} +  2 \sqrt{\eta} J^{+}_{zx}}{2 (1+\eta)}\;,\;
I^{+}_{00} =  \frac{ J^{+}_{zz} -\eta J^{+}_{xx}- 2\sqrt{\eta}
J^{+}_{zx}} {(1+\eta)} \ , \label{front1}
\end{equation}
\end{small}
with the matrix elements of the plus component of the current
evaluated  between LF polarization states denoted as $I^+_{m^\prime
m}$.

The angular condition satisfied by the matrix elements of the plus
component of the current, in the Breit-frame with $q^+=0$,  is given
by  (see e.g. \cite{Keister91}):
\begin{equation}
\Delta(q^2)= (1+2 \eta) I^{+}_{11}+I^{+}_{1-1} - \sqrt{8 \eta}
I^{+}_{10} - I^{+}_{00} =\left( J^+_{yy} - J^+_{zz} \right) \left(
1+ \eta \right) = 0 \ .\label{eq:ang}
\end{equation}
After the Eq.~(\ref{eq:ang}), the angular condition in the
IF spin basis takes a remarkable simple form:
$J^{+}_{yy}=J^{+}_{zz}$~\cite{Frankfurt93}.

We remind that, different prescriptions to extract the form factors using the
valence wave function choose three matrix elements among the four
independent ones, or any other three linearly independent
combinations of them.

{\it Evaluation of zero-mode contributions to the current.} The
contribution of the LF Z-diagram or the nonvalence contribution to
the matrix elements of the current for the vertex model
(\ref{newsymm}) is computed below.
Technically, we are able to separate out the pair terms
using the pole dislocation method, i.e., by using the limit of $q^+=\delta^+\to 0_+$
(see e.g. \cite{Pacheco98,Naus98,Pacheco992,Pacheco99}). The LF projection of
the impulse approximation formula of the current (\ref{newcu}) is
done by integration over $k^-$ in the momentum loop.

The zero-modes originated from pair terms contributions to the matrix elements
of the current in the limit $\delta^+\to 0_+$ arise from powers
of $k^-$ coming along with the Dirac trace in (\ref{newcu}). The
potential sources are the instantaneous  terms of the fermion
propagators and the derivative coupling of the fermions to the
vector particle, as in our model (\ref{newsymm}) of the vertex.
The angular condition is violated if we do not  perform the limit carefully.
We have to take into account all possible non-vanishing contributions coming
from terms of the trace carrying powers of $k^-$.

The different terms contributing to the Z-diagram in the matrix
elements of $J^+$ in the IF cartesian spin basis from the vertex
(\ref{newsymm}), consisting of $\gamma^\mu$ and derivative
couplings, are analyzed separately. They have distinct singular
behavior near the end points, which can allow zero-modes in the
limit $\delta^+\to0_+$. We compute: {\it i)} the direct term with
$\gamma^\mu$ vertices $(gg)$ from the initial and final states of
the vector particle; {\it ii)} the cross term with $\gamma^\mu$ and
derivative coupling $(dg)$; {\it iii)} the direct term with two
derivative couplings $(dd)$. This strategy simplifies the separation
of  zero-mode contributions present in the different terms of the
matrix elements of the electromagnetic current and electromagnetic
form factors of the composite vector state.

{\it i) Direct term with $\gamma^\mu$ couplings.} We start by
computing the trace with $\gamma^\mu$'s from the vertex
(\ref{newsymm}) of both the initial and final states:
\begin{equation}
Tr[gg]_{ji} =Tr[ {\sla\epsilon}^{\;\prime}_j (\sla{k}-\sla{p^\prime}
+m) \gamma^{+} (\sla{k}-\sla{p}+m) {\sla \epsilon}_i (\sla{k}+m)]\ ,
\label{trace+}
\end{equation}
where the   $k^-$ powers are separated out according to
~\cite{Pacheco992} to pin down the zero-modes. The instantaneous
term of the Dirac propagator (see e.g. \cite{Pacheco992}) brings the
$k^-$ momentum dependence in the trace:
\begin{equation}
Tr[gg]^{Z}_{ji} = \frac{1}{2} k^-\ Tr[\sla{\epsilon}^{\;\prime}_j
(\sla{k}-\sla{p^\prime} +m) \gamma^{+} (\sla{k}-\sla{p}+m)
\sla{\epsilon}_i \gamma^+ ] \ ,
\end{equation}
where the terminology ``bad'' \cite{Frankfurt79}, denoted by $Z$,
indicates that Z-diagrams or pair terms can potentially survive the
limit $\delta^+\to 0_+$ and become a zero-mode contribution to the
current. The four independent matrix elements corresponding to the
initial $(\epsilon_i)$ and final $(\epsilon^{\;\prime}_j)$
polarization states, respectively given by (\ref{inipol}) and
(\ref{finpol}), are:
\begin{equation}
Tr[ gg ]^{Z}_{xx}=  -\eta Tr[ gg ]_{zz}^{Z}, \ Tr[ gg ]^{Z}_{zx} =
-\sqrt{\eta}~Tr[ gg ]^{Z}_{zz} ,\ Tr[ gg ]^{Z}_{zz} =  ~R_{gg} ,
\label{tracegg}
\end{equation}
where $R_{gg}= \frac{k^-}{2}~Tr[ (\sla{k}-\sla{p^\prime} +m)
\gamma^{+} (\sla{k}-\sla{p}+m) \gamma^-]$ and
$Tr[ gg ]_{yy}^{Z}  =  \  4 k^- (p^+ -k^+)^2 $.

The projection over the LF hyperplane demands the integration over
$k^-$ in (\ref{newcu}). In detail, the Z-diagram contribution is:
\begin{equation}
J_{ji}^{+Z}[gg] =
 \lim_{\delta^+ \rightarrow 0_+}\;\sum_{r=4,5\; ;\;s=4,6}
\int [d^4 k]^Z \frac{ Tr[gg ]^{Z}_{ji} } {
\{1\}\{2\}\{3\}\{r\}^2\{s\}^2} \ , \label{currimgg}
\end{equation}
where $[d^4 k]^Z=[d^4 k]~\theta(p^{\prime+}-k^+) \theta(k^+-p^{+})$,
and the denominators are given by: $ \{1\} =
[k^2-m^2+\imath\epsilon]$,
 $\{2 \}  =   [(k-p)^2-m^2 + \imath \epsilon]$,
 $\{ 3 \} =   [(k-p^{\prime })^2 -  m^2 +\imath
\epsilon]$, $\{ 4 \}  =  [(k-p)^2-m_R^2 + \imath \epsilon]$, $\{5
\}= [(k-p^{\prime })^2 -  m_R^2 +\imath \epsilon]$, $\{ 6 \}  = [k^2
- m_R^2 +\imath \epsilon] $. The integration over $k^-$ is spoiled
by a end-point singularity which is taken care by using {\it the
pole dislocation method} \cite{Pacheco98}, i.e., making
$p^{\prime+}=p^++\delta^+$ in the denominator $\{3\}$. It has been
shown in \cite{Naus98} that it is enough to dislocate one of the
poles to pick  end-point singularities.

After the $k^-$ integration of (\ref{currimgg}), the Z-diagram
contributions, with support in the interval $p^+ < k^+ < p^{\prime +
}$ and in the limit $\delta^+\to0_+$, vanish for terms carrying the
dependence $(k^-)^{m+1} (p^+ -k^+)^n$ if $m < n$~\cite{Pacheco992}.
Therefore,  $J_{yy}^{+Z}=0$, and the matrix element $J_{yy}^{+}$
does not have a pair term contribution, as has been already verified
explicitly in ~\cite{Pacheco97,Pacheco992}.

Due to the trace relations (\ref{tracegg}) the zero-mode contribution to the
matrix elements of the current from the direct term with $\gamma^\mu$ couplings
are obtained from $J_{zz}^{+ Z }[gg]$ as:
\begin{equation}
J^{+Z}_{xx}[gg] =
-\eta \  J^{+Z}_{zz}[gg] ~, ~~~
J^{+Z}_{zx}[gg]  = -\sqrt{\eta} \ J^{+Z}_{zz}[gg] \ .
\label{finalgg}
\end{equation}
These relations  are the main result of this work, which are kept
valid when considering the other parts of the vector particle
vertex. The success of the prescription \cite{Inna84} in canceling
the zero-modes in the vector particle form factors is due to the
validity of (\ref{finalgg}), extending previous findings to  more
general vertex forms.

The matrix element $J_{zz}^{+ Z }[gg]$ does not vanish in the limit
of $\delta^+\to0_+$, after the $k^-$ integration in (\ref{currimgg})
is performed. In view of (\ref{finalgg}) all matrix elements of the
plus component of the current, excepting the $yy$ one, have
contribution from a zero-mode. The evaluation is performed by
dislocating the position of the zero from the denominator $\{3\}$
using $p^{\prime+}=p^++\delta^+$. The other denominators, namely
$\{i\}$ for $i\neq3$ are maintained unaltered by keeping
$p^{\prime+}=p^+$ in their expressions. Then, the Cauchy integration
in $k^-$ of (\ref{currimgg}) with $k^+$ in the interval $0< k^+-p^+
<\delta^+$, can be done by closing the contour in the upper-half of
the $k^-$ complex-plane. The arc contribution vanishes in this case.
Only the residue from the pole associated with the dislocated
denominator $\{3\}=0$, i.e.,
\begin{equation}
k^{-Z}=p^{\prime-}-\frac{({\vec p}^{\;\prime}_\perp-\vec
k_\perp)^2+m^2-\imath \epsilon}{\delta^+-(k^+-p^+)} \ ,
\label{dislocation}
\end{equation}
has to be evaluated, and then the non-vanishing zero-mode
contribution to the matrix element of the current $J_{zz}^{+ Z
}[gg]$ is obtained. Note that for $\delta^+\to 0_+$, the position of
the pole $k^{-Z}$ diverges towards infinity as $\sim 1/\delta^+$.
Therefore, terms containing powers of $k^-$ in the numerator of the
integrand are potential sources of end-point singularities, which
can produce a non vanishing result of the integration in $k^+$, even
when the interval of integration shrinks to zero, as happens in the
present case.

The above limiting behavior can be explained by taking into account
that at the pole, one has that the trace contributes with $k^-\sim
1/\delta^+$, and the denominators behave as $\{1\}\sim 1/\delta^+$
and $\{i\}\sim (\delta^+)^0$ for $i=2,4,5$. Furthermore, one has the
phase-space factor $(p^{\prime+}-k^+)\sim \delta^+$ from the
decomposition $\{3\}=(p^{\prime+}-k^+)(p^{\prime-}-k^- -
(p^{\prime}-k)^-_{on}+i\epsilon)$ where
$(p^{\prime}-k)^-_{on}=((p^\prime-k)^2_\perp+m^2)/(p^{\prime
+}-k^+)$. Putting all together, one finds that the residue is
${\mathcal O}[1/\delta^+]$ for $s=4$ in (\ref{currimgg}). Then,
after the integration in $k^+$, one has that $J_{zz}^{+ Z }[gg]$ is
finite and nonzero when $\delta^+\to 0_+$.

We add that, the LF projection of the impulse approximation formula
(\ref{newcu}) has  the valence interval $0 < k^+ <p^+ $ from the
residue at the poles $k^2=m^2$ and/or $k^2=m_R^2$ in the $k^-$
integration. These poles come from the zero of the denominators
$\{1\}$ and $\{4\}$ defined below Eq. (\ref{currimgg}).

{\it ii) Cross-term with $\gamma^\mu $ and derivative couplings.}
The trace with $\gamma^\mu$ and derivative coupling of the
constituents to the vector particle from the vertices
(\ref{newsymm}) of the initial and final states, which enters in the
impulse approximation formula for the microscopic current, reads:
\begin{equation}
Tr[dg]_{ji} = \epsilon^\prime_j\cdot (2 k
-p^\prime)\;Tr[(\sla{k}-\sla{p^\prime}+m) \gamma^{+}
(\sla{k}-\sla{p}+m) \sla{\epsilon}_i (\sla{k}+m)]
  \ , \label{trdg}
\end{equation}
where we separate out the powers in   $k^-$ to access the
zero-modes. We remind that, the Z-diagram contribution from the
$k^-$ integration in the interval $p^+ < k^+ < p^{\prime + }$ in the
limit of $\delta^+\to0_+$ vanishes for terms of the form
$(k^-)^{m+1} (p^+ -k^+)^n$ for  $m < n$. In evaluating (\ref{trdg}),
we keep only terms in the trace with $m\geq n$, which needs to be
considered for the zero-mode calculation. Under this restriction, we
get:
\begin{equation}
Tr[dg]_{ji}^Z=\epsilon^{\prime+}_j\epsilon^{+}_i
R_{dg}-4m\;k^-k^+\epsilon^{\prime+}_j\;\vec \epsilon_{i\perp}\cdot
\vec q_\perp \ , \label{trdg1}
\end{equation}
where $R_{dg}=4m\;k^-\; \left(k^-(k^+-p^+)+(\vec k_\perp-{\vec
p}^{\;\prime}_\perp)\cdot(\vec k_\perp-{\vec p}_\perp)+q_\perp\cdot
k_\perp+m^2\right)$. In the trace (\ref{trdg1}), we use
$p^+=p^{\prime^+}$, as this identification is irrelevant for the
dislocation of the pole in the denominator of the integrand in the
impulse approximation formula. Immediately, one observes that the
part of the trace, (\ref{trdg1}), which could carry a zero-mode
contribution for the $yy$ polarizations vanishes. This happens
because the $y$-polarization four-vector has plus component
identical to zero (c.f. Eqs. (\ref{inipol}) and (\ref{finpol})).

However, differently from the previous case, the analogous relations
to (\ref{tracegg}) do not hold for the trace (\ref{trdg1}). Thus,
the analysis of the $k^-$ integration should be done carefully
taking into account these traces to verify the presence of
zero-modes. In fact, the terms proportional to $k^-$ and
$(k^-)^2(k^+-p^+)$ should be checked against a zero-mode
contribution. Note that at the pole $k^{-Z}$, the product
$(k^-)^2(k^+-p^+)$ diverges like $k^-$ when $\delta^+\to0_+$. Then
for our analysis, it is enough to discuss the case when $k^-$
appears in the numerator of the integrand. As we show below, no
end-point singularity appears in this case, and the contribution to
the Z-diagram vanishes when $\delta^+\to0_+$.

The contribution of the interval $0< k^+-p^+ <\delta^+$ to the
light-front projection of the current having $\gamma^\mu$ and
derivative couplings has two possible combinations. We choose
one of them to perform our analysis without loosing generality:
\begin{equation}
J_{ji}^{+Z}[dg] =
 \lim_{\delta^+ \rightarrow 0_+}
\int [d^4 k]^Z \frac{ Tr[dg ]^{Z}_{ji} } { \{1\}\{2\}\{3\}\{6\}^2}
\left[ \frac{1}{\{4\}^2} + \frac{1}{\{5\}^2} \right] \frac{m_v} {2
(p^{\prime} \cdot k + m\; m_v-i\epsilon) }   \ , \label{gaqacurre}
\end{equation}
where the zero of  $\{3\}$ is dislocated by using
$p^{\prime+}=p^++\delta^+$, while the other denominators remain the
same. The Cauchy integration in $k^-$ with $k^+$ in the interval $0<
k^+-p^+ <\delta^+$ can be performed by closing the contour in the
upper-half of the $k^-$ complex-plane. There, two poles are present:
one from the dislocated denominator $\{3\}=0$, Eq.
(\ref{dislocation}), and another one
\begin{equation}
k^-=\frac{1}{p^{+}}\left(2\vec p_\perp^{\;\prime}\cdot \vec k_\perp
- k^+\;p^{-}-2\;m\;m_v+\imath \epsilon\right) \ . \label{poled}
\end{equation}
Next, we perform the analysis of the diverging behavior of the
several terms entering in the computation of the residue, as we have
done before. Then, we obtain that the residue from the zero of
$\{3\}$ is ${\mathcal O}[(\delta^+)^2]$, taking into account the
traces from (\ref{trdg1}). The residue from the pole (\ref{poled})
gives a contribution ${\mathcal O}[(\delta^+)^0]$. Thus, after
integration in $k^+$, one has that:
\begin{equation}J_{ji}^{+Z}[dg] \sim {\mathcal O}[\delta^+] \ .
\label{jpgd}
\end{equation}
An analogous analysis of the other possibility for the couplings
$\gamma^\mu$ and derivative of the constituents to the vector
particle in the kinematical region $0< k^+-p^+ <\delta^+$, shows
that it vanishes for $\delta^+\to 0_+$.

{\it iii) Direct term with derivative couplings.} The trace for the
case of derivative vertex couplings from (\ref{newsymm}),
corresponding to the initial and final states in the impulse
approximation formula for the microscopic current, is given by:
\begin{equation}
Tr[dd]_{ji}  =
 \left[ A_{dd}~ { k^- \over 2} +B_{dd}\right] \epsilon^\prime _j \cdot (2 k
-p^{\prime})\;\epsilon_i \cdot (2 k -p) \ . \label{trdd}
\end{equation}
 where,
\begin{eqnarray}
A_{dd} & = & Tr[(\sla{k}-\sla{p^\prime}+m) \gamma^{+}
(\sla{k}-\sla{p}+m)
\gamma^+]=8(p^+-k^+)^2 \nonumber      \\
B_{dd} & = & Tr\left[(\sla{k}-\sla{p^\prime}+m) \gamma^{+}
(\sla{k}-\sla{p}+m) \left( \frac{\gamma^-   } {2}k^+   -\vec
\gamma_{\perp}.\vec k_{\perp}+m\right) \right] \ .
\end{eqnarray}
For our analysis and with the aim of isolating terms carrying the
$k^-$ dependence, we have separated out the light-front
instantaneous terms of the Dirac propagators. However, differently
from the case {\it i)}, the analogous relations to (\ref{tracegg}),
for the trace with  $xx$, $zx$ and $zz$ polarizations do not hold
for  (\ref{trdd}).

In this case among the four possible combinations of derivative
couplings in the impulse approximation originated from the vertex, Eq.
(\ref{newsymm}), we choose to analyze, without loosing generality,
the following term: \small{
\begin{equation} J^{+Z}_{ji}[dd] =
 \lim_{\delta^+\to 0_+}\int {[d^4
k]^Z} {{m_v^2\over 4}  ~ Tr[dd]_{ji} \over
\{1\}\{2\}\{3\}\{5\}^2\{6\}^2(p \cdot k + m\; m_v -i\epsilon)
(p^\prime \cdot k + m\; m_{v}-i\epsilon)} \ , \label{intgg}
\end{equation}} where the zero of  $\{3\}$ is dislocated by using
$p^{\prime+}=p^++\delta^+$, while the other denominators are kept
unchanged. The Cauchy integration in $k^-$ is performed by closing
the contour in the upper-half of the $k^-$ complex plane. There,
three poles are present: one from the dislocated denominator
$\{3\}=0$, Eq. (\ref{dislocation}), and two others, one of them is
(\ref{poled}) and a new one:
\begin{equation}
k^-=\frac{1}{p^{+}}\left(2\vec p_\perp\cdot \vec k_\perp -
k^+\;p^{-}-2\;m\;m_v+\imath \epsilon\right) \ . \label{poled1}
\end{equation}
The residue computed at the poles (\ref{poled}) and (\ref{poled1}) are finite
and trivially do not carry an end-point singularity.
Note that the matrix element of the current with $yy$ polarizations
does not carries a zero-mode because the $k^-$ dependence in the trace appears multiplied
by $(p^+-k^+)^2$, which is enough to kill the divergence coming from the
pole $k^{-Z}$.

In the following analysis we discuss only the residue of the
integration of (\ref{intgg}) in $k^-$ at the pole $k^{-Z}$
(\ref{dislocation}). Because, this residue can potentially give a
zero-mode contribution to the matrix elements of the current picking
up end-point singularities. The scalar product in (\ref{trdd})
computed for the cartesian polarization four-vectors $x$ and $z$
given by Eqs. (\ref{inipol}) and (\ref{finpol}), provide terms up to
$(k^-)^2$, and then Eq. (\ref{trdd})  carries terms in $k^-$ up the
third power. However, it is sufficient to analyze terms proportional
to $(k^-)^2$ from (\ref{trdd}), because terms with $(k^-)^3$ come
multiplied by $(k^+-p^+)^2$ in this case. The product
$(k^{-Z})^3(k^+-p^+)^2$ behave as $k^{-Z}$ when approaching the
end-point of the $k^+$ integral in (\ref{intgg}). Considering the
denominators and $(k^-)^2$ in the numerator of (\ref{intgg}), the
residue at $k^{-Z}$ is ${\mathcal O}[(\delta^+)^2]$.  Thus, after
integrating in $k^+$, one has that:
\begin{equation}
J^{+Z}_{ij}[dd]   \sim \ {\mathcal O}[\delta^{+}]  \ .
\end{equation} Evoking the same analysis as we did above, the other
terms coming from derivative couplings in the kinematical region $0<
k^+-p^+ <\delta^+$ vanish when $\delta^+$ goes to zero.

{\it Cancelation of zero-modes in the electromagnetic form factors.}
The contributions from Z-diagrams survive the limit $q^+\to0_+$ in
the matrix elements of $J^+$, only for the direct term with
$\gamma^\mu$ couplings. The relations (\ref{finalgg}) are valid for
the full matrix elements of the current, because, as we have shown,
the zero-modes vanish for the other possible combinations from the
vertices of the initial and final states of the vector particle.
Then, for the full $J_{zz}^{+ Z } \sim {\mathcal O}[(\delta^+)^0]$,
the relations
\begin{equation}
J^{+Z}_{xx} +\eta \  J^{+Z}_{zz}=0 \ ,    J^{+Z}_{zx} +\sqrt{\eta} \
J^{+Z}_{zz}=0\ ~~\text{and}~~ J^{+Z}_{yy}=0, \label{finalcurr}
\end{equation}
are satisfied. From the angular condition, $J^+_{zz}=J^{+}_{yy}$,
and considering that the $yy$ matrix element does not carry a
zero-mode, one can express the non vanishing zero-mode matrix
element in terms of the matrix elements computed in the valence
region as $J^{+Z}_{zz}=J^{+V}_{yy}-J^{+V}_{zz}$ (the superscript $V$
indicates the valence terms). Then, we have that:
\begin{equation} J^{+}_{xx}=
J^{+V}_{xx}-\eta \left(J^{+V}_{yy}-J^{+V}_{zz}\right) , \ J^{+}_{zx}
=J^{+V}_{zx}-\sqrt{\eta}\left(J^{+V}_{yy}-J^{+V}_{zz}\right) , \
J^+_{zz}=J^{+V}_{yy} ,\label{zrcurr}
\end{equation}
which isolates the zero-modes in the  matrix elements and can be
computed considering only the valence region.

In particular, introducing the relations (\ref{finalcurr}) in
(\ref{front1}), one obtain the corresponding to matrix elements in
the LF spin basis as:
\begin{equation}
I^{+Z}_{11} = 0\;,\ I^{+Z}_{10} = 0\;,\ I^{+Z}_{1-1}= 0\ \text{and}
\ I^{+Z}_{00} = (1+\eta )J_{zz}^{+Z} \ , \label{eq:ifront1}
\end{equation}
with $\lim_{\delta^+\to0_+}J_{zz}^{+Z} \neq 0$, which gives a
zero-mode contribution only in this case.

Within the prescription of Grach and Kondratyuk (GK) \cite{Inna84}
the matrix element $I^{+}_{00}$ is eliminated from the form factors.
It trivially excludes the zero-modes due to the validity of
(\ref{eq:ifront1}). To appreciate this finding in the IF spin basis,
we write below the form factors,
\begin{multline} G_0^{GK} = \frac{1}{3} \left( J_{xx}^{+} + \eta
J_{zz}^{+}+ (2- \eta)J_{yy}^{+}\right) , \\ G_1^{GK} = J^+_{yy}
-\frac{1}{\sqrt{\eta}}\left( J_{zx}^{+}+\sqrt{\eta}J_{zz}^{+}
\right) ,
\\
G_2^{GK}  =  \frac{\sqrt{2}}{3} \left( J_{xx}^{+}+ \eta J_{zz}^{+} -
(1+ \eta)J_{yy}^{+}  \right) , \label{ffactors}
\end{multline}
by  transforming the matrix elements from the LF spin
basis~\cite{Inna84} to the IF spin basis \cite{Frankfurt93}. One
immediately recognizes from (\ref{ffactors}) and the relations
(\ref{finalcurr}) for the contributions of the zero-mode the matrix
elements in the IF spin basis that
\begin{equation} G_0^{GK,\;Z} =G_1^{GK,\;Z}  =
G_2^{GK,\;Z}  = 0 , \label{fiffactor}
\end{equation}
with form factors (\ref{ffactors}) giving by considering only the
valence region. Note that, if the relations (\ref{zrcurr}) are taken
into account, the computation of the form factors are independent of
the prescription chosen.

The above findings in Eq. (\ref{fiffactor})  generalizes the
conclusion of \cite{JI2002} obtained with a smeared photon vertex
model to the symmetrical vector meson vertex model, where only the
valence region contributes to the form factors in the GK
prescription. Also found in a numerical calculation of the
$\rho$-meson form factors by comparing the LF calculation
considering only the valence region with the covariant results of
the model \cite{Pacheco97}.

{\it Summary.} We analyzed the rotational symmetry properties of the
matrix elements of the plus component of the electromagnetic current
in the Breit-frame with $q^+=0$ for a symmetric and analytic model
of a spin-1 composite particle vertex, considering the projection
onto the light-front. If only the valence region is computed in the
impulse approximation formula, rotations are not properly accounted
by the matrix elements, and the angular condition is violated. This
is why different prescriptions for extracting form factors from the
microscopic matrix elements do not provide an unique answer, which
led to alternative proposals to calculate form factors like, e.g.,
the Lev-Pace-Salm\`e frame ($\vec q_\perp=0$).

The naive computation of the microscopic matrix elements of $J^+$
for $q^+=0$,  relying only on the valence region, leads to the
violation of the angular condition even in analytical models
\cite{Pacheco97} and \cite{JI2002} of composite spin-1 particles.
Here, we have used a more general form of the spin-1 vertex to
compute the matrix elements of $J^+$ using the instant-form
polarization basis, in the  limit of $q^+\rightarrow 0_+$. We showed
how to single out the contribution of zero-modes making use of the
limit $q^+\rightarrow 0_+$. We prove that the prescription suggested
by Grach and Kondratyuk to extract the form factors from the
microscopic current, which excludes in the light-front helicity
basis the $0\to 0$ matrix element of $J^+$ among the four
independent ones, eliminates unwanted zero-modes, keeping
contribution only from the valence region. Our derivation
generalizes to symmetric models the above conclusion found for a
simplified non-symmetrical form of the the $\rho$-meson vertex with
point-like quarks and also for a model of a smeared quark-photon
vertex. Our methods are suitable for applications, e.g., to study
vector meson elastic form factors, and also  can be easily  extended
to study transition form factors involving spin-1 composite
particles.

\nl {\bf Acknowledgments.} This work was supported in part by the
Brazilian agencies FAPESP (Funda\c{c}\~ao de Amparo \`a Pesquisa do
Estado de S\~ao Paulo) and CNPq (Conselho Nacional de
Desenvolvimento Cient\'\i fico e Tecnol\'ogico).

\end{document}